%% file: main.tex
\documentclass[conference]{IEEEtran}
\AtBeginDocument{%
  }

\usepackage{graphicx} 
\usepackage{xspace}
\usepackage{tabularray}
\usepackage{tcolorbox}
\usepackage{multirow}
\usepackage{booktabs}
\usepackage{pgfplots}
\usepackage{pgfplotstable}
\usepackage{adjustbox}
\usepackage{subcaption}
\usepackage{adjustbox}
\usepackage{colortbl}
\usepackage{tcolorbox}
\usepackage{listings}
\usepackage{algorithm}
\usepackage{algorithmic}
\usepackage{amsfonts}
\usepackage{longtable} 

\input{macros}

\title{Enhancing LLM-Based Code Generation with Complexity Metrics: A Feedback-Driven Approach}

\author{\IEEEauthorblockN{Melika Sepidband}
\IEEEauthorblockA{\textit{Lassonde School of Eng.} \\
\textit{York University}\\
Toronto, Canada\\
melikasp@yorku.ca}
\and
\IEEEauthorblockN{Hamed Taherkhani}
\IEEEauthorblockA{\textit{Lassonde School of Eng.} \\
\textit{York University}\\
Toronto, Canada\\
hamedth@yorku.ca}
\and
\IEEEauthorblockN{Song Wang}
\IEEEauthorblockA{\textit{Lassonde School of Eng.} \\
\textit{York University}\\
Toronto, Canada\\
wangsong@yorku.ca}
\and
\IEEEauthorblockN{Hadi Hemmati}
\IEEEauthorblockA{\textit{Lassonde School of Eng.} \\
\textit{York University}\\
Toronto, Canada\\
hemmati@yorku.ca}
}

\begin{document}

\maketitle

\begin{abstract}
    \input{sec/0_abstract}
\end{abstract}
\input{sec/1_introduction}

\input{sec/2_RelatedWork}\
\input{sec/3_empirical_study}
\input{sec/6_Results}
\input{sec/7_Threats}
\input{sec/8_Conclusion}

\balance
\bibliographystyle{IEEEtran}
\bibliography{refs}
\end{document}

%% file: macros.tex
\usepackage{soul}
\usepackage{hyperref}
\usepackage{multirow}
\usepackage{listings}
\usepackage{fancybox}
\usepackage{enumitem}
\usepackage{graphicx}
\usepackage{color}
\usepackage{array}
\usepackage{amsmath}
\usepackage{xspace}
\usepackage{url}
\usepackage{tikz}
\usepackage{caption}
\usepackage{pgfplots}
\usepackage{balance}
\usepackage{subcaption}
\pgfplotsset{compat=1.16}
\usepackage{pgf-pie}
\usetikzlibrary{pgfplots.statistics,calc}
\usepackage{algorithm}
\usepackage{algorithmic}
\usepackage{adjustbox}
\usepackage{makecell}

\usepackage{tikz}
\usepackage{calc}

\usepackage{tcolorbox}
\makeatletter
\newcommand{\mybox}[1]{%
	\setbox0=\hbox{#1}%
	\setlength{\@tempdima}{\dimexpr\wd0+13pt}%
	\begin{tcolorbox}[boxrule=0.5pt, colback=white, arc=4pt,
		left=6pt,right=6pt,top=6pt,bottom=6pt,boxsep=0pt]
		#1
	\end{tcolorbox}
}

\definecolor{songcolor}{RGB}{191,191,191}



%% file: sec/0_abstract.tex
Automatic code generation has gained significant momentum with the advent of Large Language Models (LLMs) such as GPT-4. Although many studies focus on improving the effectiveness of LLMs for code generation, very limited work tries to understand the generated code's characteristics and leverage that to improve failed cases. 
In this paper, as the most straightforward characteristic of code, we investigate the relationship between code complexity and the success of LLM-generated code. 
Using a large set of standard complexity metrics, we first conduct an empirical analysis to explore their correlation with LLM's performance on code generation (i.e., Pass@1). 
Using logistic regression models, we identify which complexity metrics are most predictive of code correctness.  
Building on these findings, we propose an iterative feedback method, where LLMs are prompted to generate correct code based on complexity metrics from previous failed outputs. 
We validate our approach across multiple benchmarks (i.e., HumanEval, MBPP, LeetCode, and BigCodeBench) and various LLMs (i.e., GPT-4o, GPT-3.5 Turbo, Llama 3.1, and GPT-o3 mini), comparing the results with two baseline methods: (a) zero-shot generation, and (b) iterative execution-based feedback without our code complexity insights. 
Experiment results show that our approach makes notable improvements, particularly with a smaller LLM (GPT-3.5 Turbo), where, e.g., Pass@1 increased by 35.71\% compared to the baseline's improvement of 12.5\% on the HumanEval dataset. 
The study expands experiments to BigCodeBench and integrates the method with the Reflexion code generation agent, leading to Pass@1 improvements of 20\% (GPT-4o) and 23.07\% (GPT-o3 mini). The results highlight that complexity-aware feedback enhances both direct LLM prompting and agent-based workflows.

%% file: sec/1_introduction.tex
\section{Introduction}



Automatic code generation aims to reduce manual coding and boost productivity~\cite{zhang2024no}, with LLMs like GPT-4~\cite{achiam2023gpt} making significant advancements. However, ensuring accuracy and correctness remains a challenge.
Recently, several approaches have been proposed to enhance LLM-based code generation. These include prompt engineering techniques like chain-of-thought reasoning~\cite{wei2022chain}, which encourages the model to break problems step by step. Additionally, feedback-based methods have shown promise in correcting generated code. For instance, Reflexion~\cite{shinn2023reflexion} helps language agents learn from mistakes by using feedback and storing reflections in memory, improving their decisions without changing their underlying model.

Most feedback-based approaches only take into account the execution results. In this paper, we explore another characteristic of the generated code, which is its complexity. The idea is that by understanding the expected level of complexity (defined by various metrics) for the correct code of a programming task in natural language, we can effectively guide LLMs toward generating the correct code (with the right complexity), even when the initial prompt results in an incorrect solution. 
Additionally, we investigate whether this complexity-aware method can enhance the performance of existing feedback-based approaches, further refining code generation outcomes.

Common complexity metrics that can be useful predictors of correct vs incorrect code may include cyclomatic complexity~\cite{ebert2016cyclomatic}, which measures the number of independent paths through a program's control flow, and Halstead complexity~\cite{hariprasad2017software}, which evaluates the size and volume of code based on operators and operands. 
In this work, we have employed 53 widely-used complexity metrics (details are in Section~\ref{sec:comp}) 
as a quantifiable measure of the complexity inherent in the generated code.


%

As a motivation example, consider Figure \ref{fig:compare}, which presents a solution generated by GPT-4o~\cite{openai2024gpt4o} on an example from the HumanEval~\cite{chen2021evaluating} dataset. The prompt for this task is \textit{``Given two positive integers a and b, return the even digits between a and b, in ascending order''}.
The initial incorrect code from GPT-4o (Listing 1) produces even numbers within {[a, b]} but includes numbers beyond the specific digits required \textit{\{2, 4, 6, and 8\}}, leading to an overgeneralized solution. By contrast, the correct code generated by our approach (Listing 2) limits the output to only the specified even digits by checking for membership in \textit{\{2, 4, 6, 8\}}, aligning closely with the problem requirements. 
Looking at the complexity metrics associated with these listings, we can see that, for instance, the Halstead effort metric is 94.89, 48.60, and 42.79 in Listing 1, Listing 2, and Listing 3, respectively. This shows that some complexity metrics can be predictors of code correctness. 


\lstset{
    language=Python,
    basicstyle=\footnotesize, 
    keywordstyle=\color{blue}\bfseries, 
    commentstyle=\color{teal}\itshape, 
    stringstyle=\color{red}, 
    numberstyle=\tiny\color{gray}, 
    stepnumber=1,
    numbersep=10pt, 
    breaklines=true, 
    frame=tb, 
    framerule=0.5pt, 
    rulecolor=\color{black},
    backgroundcolor=\color{lightgray!20}, 
    tabsize=4, 
    showstringspaces=false, 
    captionpos=b, 
}

\begin{figure*}[ht]
    \centering
    \begin{minipage}{0.32\textwidth}
        \begin{lstlisting}[caption={\small{code generated by the LLM}}]
def generate_integers(a, b):
    if a > b:
        a, b = b, a
    even_numbers = [i for i in range(a, b + 1) if i % 2 == 0]
    
    return even_numbers
\end{lstlisting}

    \end{minipage}
    \hfill
    \begin{minipage}{0.32\textwidth}
        \begin{lstlisting}[caption={\small{code generated by our approach}}]
def generate_integers(a, b):
    if a > b:
        a, b = b, a
    even_digits = []
    for i in range(a, b + 1):
        if i in {2, 4, 6, 8}:
            even_digits.append(i)
    return even_digits
\end{lstlisting}
    \end{minipage}
    \hfill
    \begin{minipage}{0.32\textwidth}
        \begin{lstlisting}[caption={\small{ground truth solution}}]
def generate_integers(a, b):
   
    lower = max(2, min(a, b))
    
    upper = min(8, max(a, b))
    
    return [i for i in range(lower, upper+1) if i % 2 == 0]
\end{lstlisting}
    \end{minipage}
    \caption{Comparison of incorrect code generated by the LLM (left), the correct code generated by our approach (middle), and the ground truth code in the dataset (right). The incorrect code shows higher complexity based on certain metrics, such as Halstead metrics, the number of numeric literals, and the frequency of mathematical operations. For instance, the Halstead effort metric is 94.89 in the left code and 48.60 and 42.79 in the middle and right code, respectively.}
    \label{fig:compare}
\end{figure*}

To investigate the correlation between code complexity and LLM's effectiveness and its potential to improve LLMs for code generation using the code complexity feedback, we conduct a study with these four research questions:

\textbf{RQ1: Are complexity metrics of the generated codes correlated with the code generation's effectiveness (pass@1)?}\\
Our first objective is to investigate whether there is a correlation between the complexity metrics of the generated code and the success rate of the LLMs' outputs, measured as Pass@1~\cite{chen2021evaluating} (i.e., the percentage of correct solutions on the first attempt). Using a machine learning model—specifically logistic regression~\cite{lavalley2008logistic}—We observe a clear correlation between Pass@1 and specific complexity metrics. This relationship is particularly strong in the HumanEval dataset, where models such as GPT-4o attain very high accuracy. 

\textbf{RQ2: How do the distributions of complexity metrics differ between successful and failed code solutions generated by LLMs?}\\
We analyze the distribution of each metric across both correct and incorrect code samples generated by LLMs, which enabled us to identify patterns and differences in metric values associated with successful versus unsuccessful code generations. To gain deeper insights, we divide this research question into two sub-questions: \\
\textbf{RQ2.1: Are there specific metrics that LLMs have more difficulty getting right when generating code?} This sub-question investigates which metrics are harder for different LLMs, such as GPT-4o, GPT-3.5 Turbo~\cite{gpt3.5-turbo}, and Llama 3.1~\cite{dubey2024llama}, to optimize, examining success and failure cases to highlight LLM-specific challenges.\\
\textbf{RQ2.2: Do different datasets' characteristics differ in terms of code complexity metrics' distributions over the correct vs. incorrect code?} 
Here, we assess whether complexity metrics influence success differently across datasets (HumanEval, MBPP~\cite{austin2021program}, LeetCode\cite{leetcode}), providing insights into dataset-specific characteristics.

\textbf{RQ3: Can feedback based on complexity metric values of the generated code improve LLMs' code generation effectiveness?}

Having established this correlation, we sought to leverage these complexity metrics to further help LLMs refine the generated code. Using a diverse set of datasets and different LLMs, we conducted experiments to refine code generation iteratively. By calculating the Shapley values~\cite{roth1988introduction} of the complexity metrics, we identified the most important metrics for each dataset and used them as feedback to prompt the LLM to generate new code with different complexity characteristics.

To prevent overfitting, new test cases are generated using GPT-4o, and if the code fails, we identify the five most impactful complexity metrics, prompting the LLM to regenerate and alter those metrics in the generated code. This cycle continues until the code passes all test cases or reaches a maximum number of iterations (five in our study).


\textbf{RQ4: Can feedback based on complexity metric values enhance the effectiveness of code generation agents, particularly on datasets with lower accuracy?}

Here, we explore whether our complexity-based feedback method can further improve code generation when integrated into an agent-based framework. Given that datasets like HumanEval, MBPP, and LeetCode already have high Pass@1 scores, we focus on a more challenging dataset—BigCodeBench~\cite{zhuo2024bigcodebench}—where there is greater room for improvement. We apply our complexity-aware feedback method on top of a code generation agent (Reflexion~\cite{shinn2023reflexion}) to assess its effectiveness in refining agent-generated code. Our results show that even in more complex scenarios, guiding the agent with complexity-based insights leads to measurable improvements in code accuracy.

In short, the contributions of this paper are:
\begin{enumerate}
    \item Demonstrating the correlation between code complexity metrics and LLMs' code generation success (Pass@1).
    \item Introducing an iterative feedback method based on code complexity metrics to enhance the correctness of generated code, both in standalone LLM prompting and within an agent-based framework.
    \item Conducting comprehensive experiments across multiple datasets and LLMs to validate our approach.
\end{enumerate}

\noindent \textbf{Data Availability:} 
We release the source code of our experiments to help other researchers replicate and extend our study\footnote{https://github.com/MelikaSepidband/Code-Generation-Complexity-Metrics/}.


%% file: sec/2_RelatedWork.tex
\section{Background and Related Works}

\subsection{LLMs for Code Generation}


Recent LLMs have brought significant improvements to code generation. Codex~\cite{chen2021evaluating}, based on GPT-3 and trained on extensive code repositories, was developed specifically for code tasks and stands out in both comprehension and generation. CodeLlama~\cite{roziere2023code}, an enhanced variant of Llama 2~\cite{touvron2023llama} specifically fine-tuned for coding tasks, excels in code generation, and is offered in multiple parameter sizes. Llama 3 is a powerful successor to Llama 2, available in various parameter configurations, making it an excellent tool for generating code efficiently.

In this research, we use GPT-4o, a high-performing, closed-source model, GPT-o3 mini~\cite{gpt-o3-mini}, OpenAI's most cost-effective model in its reasoning series, alongside GPT-3.5-turbo, another advanced closed-source model, and Llama 3.1, a cost-effective, open-source model. GPT-4o and GPT-o3 mini are selected for their superior capabilities in closed-source applications, while GPT-3.5-turbo offers a balance of performance and efficiency in various coding tasks. Llama 3.1 provides a budget-friendly and accessible alternative for open-source experiments, complementing the other models in our study.

\subsection{Feedback-Based Code Generation} 

Recent frameworks like Parsel~\cite{zelikman2023parsel} and ANPL~\cite{huang2023anpl} enhance code generation by structuring tasks and refining code iteratively. Other systems, such as DyLAN~\cite{liu2023dynamic}, Reflexion~\cite{shinn2023reflexion}, and AgentCoder~\cite{huang2023agentcoder}, leverage multi-agent and feedback-based methods to improve task efficiency. The LDB~\cite{zhong2024ldb} framework breaks down code for error detection, while EPiC~\cite{taherkhani2024epic} uses evolutionary algorithms to optimize prompts for efficient code generation.

Our approach uniquely leverages the complexity metrics of generated code as feedback to refine outputs, representing an innovative feedback-based method focused on improving code quality through metric-based insights.

\subsection{Code Complexity Metrics} 
Software complexity metrics are vital for managing quality and reducing costs. Yu and Zhou~\cite{yu2010survey} provide a comprehensive review, highlighting their role in improving maintainability throughout the software lifecycle.

Zamani and Hemmati~\cite{zamani2021pragmatic} introduce “Tuning Gain,” a metric estimating cost-effectiveness in search-based test generation. Using metrics like McCabe’s and Halstead’s complexities, they demonstrate the value of static metrics in enhancing software testing. 
Mashhadi et al.~\cite{mashhadi2024empirical} assess code complexity metrics for bug prediction. They find Lines of Code and McCabe’s complexity effective for bug detection but insufficient for severity prediction, suggesting a need for context-sensitive metrics. 
Harzevili and Alizadeh~\cite{harzevili2021analysis} discuss complexities in software defect prediction, noting that traditional classifiers often overlook interdependencies among metrics, which can impact real-world applications.

Despite the wealth of research on complexity metrics, no prior studies have focused on using these metrics specifically to improve the quality of code generated by LLMs. In this paper, we aim to fill this gap by applying complexity metrics to enhance the code generation process.

%% file: sec/3_empirical_study.tex
\section{Study Setup}

In this section, we explain the datasets, models, complexity metrics, evaluation metrics, and experiment setup and design for our study. 

\subsection{Datasets}
\label{sec:datasets}

For our experiments, we used four datasets: HumanEval~\cite{chen2021evaluating}, MBPP-sanitized~\cite{austin2021program}, LeetCode~\cite{leetcode}, and BigCodeBench~\cite{zhuo2024bigcodebench}. 
\textbf{HumanEval} consists of 164 Python programming problems, each with a function signature, description, and examples, widely used to evaluate the accuracy of LLM code generation. \\
\textbf{MBPP-sanitized} is a refined version of MBPP, addressing inconsistencies in prompts and test cases for improved reliability. It includes 120 training samples and 257 test samples. \\
\textbf{LeetCode} is a large dataset of 2,360 Python problems commonly used in competitive programming. We focused on 561 problems with verified correct code, extracting test cases embedded in problem descriptions. \\
\textbf{BigCodeBench} assesses LLMs on practical and complex coding tasks. We used BigCodeBench-Hard, a 148-task subset, featuring more intricate instructions and diverse function calls.

\subsection{Models}
For code generation, we experimented with a range of LLMs to capture the differences in their capabilities, i.e., GPT-4o~\cite{openai2024gpt4o}, GPT-3.5 Turbo~\cite{gpt3.5-turbo}, Llama 3.1~\cite{dubey2024llama},  and GPT-o3 mini~\cite{gpt-o3-mini}.





\subsection{Complexity Metrics}
\label{sec:comp}

Complexity metrics are quantitative measures used to assess various aspects of code, such as its structure, readability, and maintainability. These metrics help to understand how complex a piece of code is. For our study, we used a comprehensive set of 53 complexity metrics from the literature~\cite{wang2016automatically,wang2021continuous}, which are explained in Table \ref{tab:complexity_metrics}.

\begin{table}[ht]
    \centering
    \rowcolors{2}{gray!15}{white} 
    \footnotesize
     \caption{Overview of Code Complexity Metrics Used in our Study}
    \label{tab:complexity_metrics}
    \begin{tabular}{>{\bfseries}p{2.5cm} p{5.5cm}} 
        \toprule
        Complexity Metric & Description \\
        \midrule
        Cyclomatic Complexity & Measures the number of linearly independent paths through the code, indicating its control flow complexity. \\
        Halstead Complexity & Encompasses sub-metrics: \emph{Length (N)}: total count of operators and operands; \emph{Vocabulary (n)}: unique operators and operands; \emph{Volume (V)}: \(V = N \log_2(n)\); \emph{Difficulty (D)}: \(D = \frac{n_1}{2} \times \frac{N_2}{n_2}\); \emph{Effort (E)}: \(E = D \times V\); and \emph{Time (T)}: estimated development time. \\
        Maintainability Index & A composite metric combining Cyclomatic Complexity, Halstead Volume, and lines of code to assess overall maintainability. \\
        Number of Lines & Total count of lines in the source code. \\
        Number of Loops & Count of loop constructs (e.g., \texttt{for}, \texttt{while}). \\
        Number of Comparisons & Count of conditional comparisons (e.g., \texttt{==}, \texttt{!=}). \\
        Number of Variables & Count of declared and utilized variables. \\
        Number of String Literals & Total count of string literals present in the code. \\
        Number of Numeric Literals & Total count of numeric literals present in the code. \\
        Number of Math Operations & Count of arithmetic operations (e.g., \(+\), \(-\), \(*\), \(/\), \(\gg\), \(\ll\)). \\
        Maximum Nested Blocks & Maximum depth of nested control structures in the code. \\
        Number of Unique Words & Count of distinct tokens or words in the code. \\
        Number of Each Python Keyword & Frequency count of each of the 35 Python keywords (e.g., \texttt{if}, \texttt{for}, \texttt{return}). \\
        \bottomrule
    \end{tabular}
   
\end{table}

\subsection{Evaluation Metric}

In this work, we use pass@k~\cite{chen2021evaluating} to measure the performance of LLMs in code generation tasks. 
Pass@k calculates the probability that at least one of the k-generated code outputs passes all the test cases for a given coding problem. 
Pass@1 is a specific case of the pass@k metric in which only the top-ranked generated solution is considered. It measures the probability that the first generated code sample passes all test cases, assessing the model's ability to produce a correct solution on the first attempt. In this study, pass@1 is particularly crucial, as it directly measures the initial success rate of LLMs in producing correct codes. A higher pass@1 score indicates better code generation performance.

\subsection{Experimental Setup}

\subsubsection{Preprocessing}
We extracted complexity metrics from the generated code and ground truth for all experiments. Cyclomatic and Halstead complexities were computed using the Radon library in Python, while custom functions were defined for other complexity metrics. To normalize these features, we applied StandardScaler to ensure all metrics were on the same scale for the prediction process. 

\subsubsection{Configurations of LLMs}

GPT-3.5-turbo, GPT-4o, and GPT-o3 mini were accessed via Colab with high RAM configurations on the CPU. 
Meta-Llama-3.1-8B-Instruct was run using an A100 GPU for increased speed.
The models were configured with a temperature of 0.2, max\_tokens set to 1000, and a frequency\_penalty of 0.0 to control creativity and output length. We used this prompt to generate code based on the function description:

\begin{tcolorbox}[colback=gray!10, colframe=red!10!black, arc=0mm, auto outer arc, boxsep=2pt, left=2pt, right=2pt, top=2pt, bottom=2pt]
Please complete the Python function:\{function\_description\}
\end{tcolorbox}



    
In cases where the generated code was incorrect, feedback was provided to the LLM using complexity metrics: 

\begin{tcolorbox}[colback=gray!10, colframe=red!10!black, arc=0mm, auto outer arc, boxsep=2pt, left=2pt, right=2pt, top=2pt, bottom=2pt]
The previously generated code is incorrect. Please complete the Python function according to the feedback: \{feedback\}
\end{tcolorbox}

Where feedback took the form:

\begin{tcolorbox}[colback=gray!10, colframe=red!10!black, arc=0mm, auto outer arc, boxsep=2pt, left=2pt, right=2pt, top=2pt, bottom=2pt]
Please ensure that your generated code has different values for the following complexity metrics: \{metrics\}
\end{tcolorbox}

\subsubsection{Logistic Regression Model}

To predict the success of the generated code (pass@1), we trained a Logistic Regression model on the complexity metrics of the generated code. We used max\_iter=10000 and penalty='l2' to handle feature regularization and to ensure convergence for all training instances.

\subsection{Experiment Design}

\subsubsection{RQ1 Design}
In this RQ, we first prompt LLMs to generate code solutions for problems in each dataset. For each code solution, we compute various complexity metrics (as discussed in Section~\ref{sec:comp}). 
Next, we apply a Logistic Regression model with 5-fold cross-validation to analyze the correlation between metric values and the likelihood of code solutions' success (pass@1).

This step involved using complexity metrics as predictor variables to assess their ability to predict the success or failure of generated code. The goal is to understand whether higher or lower values of specific metrics are associated with successful outcomes. To enhance the model's accuracy, we applied several feature selection methods, including:

    \textbf{L1 Regularization (Lasso)~\cite{ng2004feature}: } This method reduces model complexity by setting coefficients of less relevant features to zero, focusing on key complexity metrics.
    
    \textbf{Recursive Feature Elimination (RFE)~\cite{guyon2002gene}: } This method iteratively removes the least important features based on model performance to identify a minimal, effective subset for predicting pass@1.
    
    \textbf{Correlation-based Selection~\cite{hall1999correlation}: } This method chooses features strongly correlated with the target (pass@1) and removes redundant ones to ensure each selected feature uniquely contributes to prediction.
    
    \textbf{Shapley Values~\cite{lundberg2017unified}:} This method calculates each feature's contribution to a model by averaging its impact across all feature combinations. For feature selection, it ranks features by their Shapley values, helping identify and retain the most influential ones, which enhances model interpretability.
    
 \subsubsection{RQ2 Design}
To explore how complexity metric values differ between successful (pass@1 = 1) and failed (pass@1 = 0) instances, we analyze the distribution of metrics for each case. We identify for which metrics the difference between the correct vs. incorrect instances is significant. 
This analysis reveals insights into which complexity metrics are more often associated with failures and thus are critical for LLMs to get them right and to produce correct code.
We then extend our analysis, and for each metric, we compare successful and failed cases across GPT-3.5-turbo, GPT-4, and Llama 3.1-8B-Instruct. Additionally, to detect any dataset-specific trends, we examine whether different datasets show varying distributions of complexity metrics between successful and unsuccessful solutions by comparing their median metric values.


\subsubsection{RQ3 Design}
\begin{figure*}
    \centering
    \includegraphics[width=1.0\linewidth]{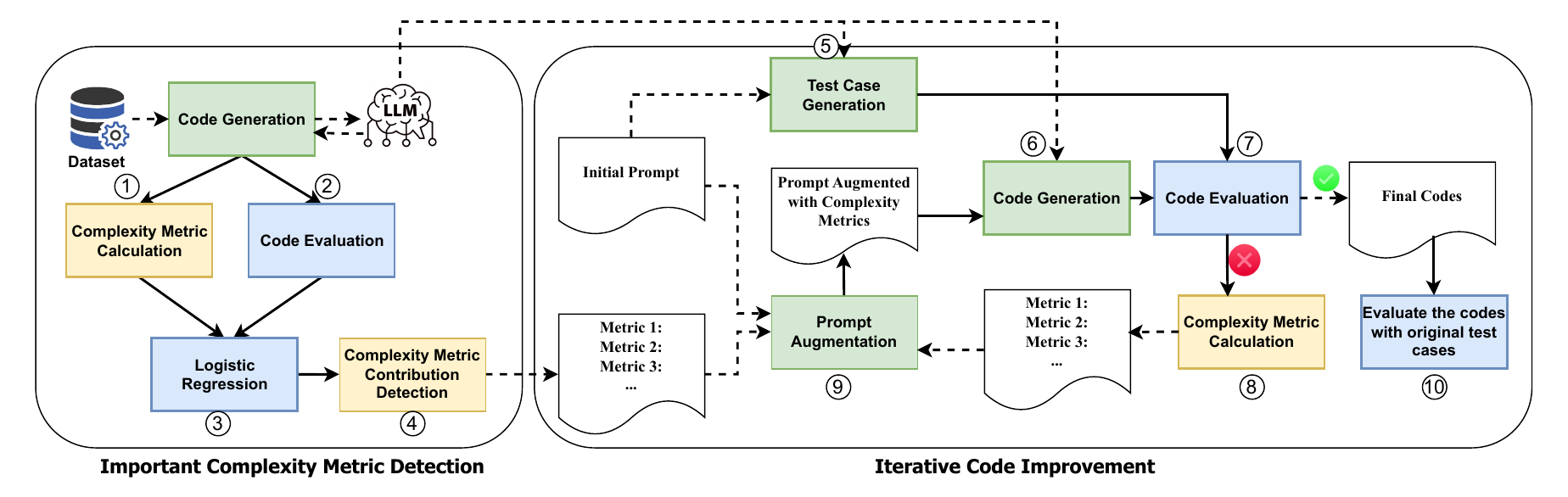}
    \caption{Overview of Complexity-Aware Feedback for Enhanced LLM Code Generation}
    \label{fig:fig1}
\end{figure*}

RQ3 is about assessing our proposed feedback-driven mechanism to enhance the code generation process. Figure \ref{fig:fig1} provides an overview of our method, which will be explained in detail in the rest of this section. Our method is divided into two parts: \textbf{Important Complexity Metric Detection} and \textbf{Iterative Code Improvement}.

\textbf{Important Complexity Metric Detection:} 
The goal of this phase is to identify the most predictive complexity metrics per dataset. We used the training set of each dataset to prompt the LLM for code generation. For HumanEval, LeetCode, and BigCodeBench, this involves using 4 folds in the cross-validation process. For MBPP, we used the provided training set. In \textbf{Step 1}, we compute the complexity metrics for each generated code. In \textbf{Step 2}, we evaluate their correctness using the pass@1 criterion. In \textbf{Step 3}, we employ a Logistic Regression model to predict the pass@1 using the metrics as the feature set and pass@1 as labels in the training set. In \textbf{Step 4}, the feature importance of the trained model is analyzed using Shapley values~\cite{lundberg2017unified}. Shapley values provide a robust approach for assessing the contribution of each feature to the model's prediction. Here, Shapley values were used to measure the impact of various complexity metrics on the likelihood of a generated code passing all test cases (i.e., pass@1). By assigning importance scores to each metric, Shapley values enabled a nuanced understanding of how individual complexity characteristics influence model predictions, highlighting which metrics are critical to predicting code correctness.

\textbf{Iterative Code Improvement:} 
In this phase, we apply an iterative process for code improvement on the evaluation set (the remaining fold in cross-validation in HumanEval, LeetCode, and BigCodeBench or the evaluation set in MBPP). The process starts with \textbf{Step 5}, where test cases are generated using GPT-4o for internal code evaluation. The algorithm utilizes test cases generated by GPT-4o instead of the dataset's evaluation test cases, as it continues processing only the samples that fail these test cases. Incorporating ground truth tests within this feedback loop would introduce data leakage. Next, in \textbf{Step 6}, we prompt the LLM to generate new code. We then evaluate the generated code (\textbf{Step 7}), and if a code sample passed all generated test cases, it was added to the final code pool. In \textbf{Step 8}, for any code that failed, the values of the five most influential metrics (identified in \textbf{Step 4}) are collected. 
This feedback will be provided to LLMs to adjust these complexity aspects in the regenerated code, in \textbf{Step 9}, with the following prompt: ``\textit{Please ensure that your generated code has different values for the following complexity metrics: \{metrics\}}'' where \textit{metrics} are the five most influential metrics identified earlier. 
This refinement process will be repeated over a maximum of $N$ iterations ($N=5$ in this study) or until the generated code successfully passes all test cases. 
Lastly, in \textbf{Step 10},  the final set of generated code samples will be evaluated using the original test cases in the dataset (the developer-written ones and not the LLM-generated tests, which are used internally in the algorithm).

This process is also formally presented as Algorithm \ref{alg:alg1}, which comprises the above two phases. Lines 1 through 7 correspond to the process illustrated as "Important Complexity Metric Detection" in Figure \ref{fig:fig1} (training set) and focus on identifying the most impactful complexity metrics. Lines 8 onward relate to the feedback and evaluation phase, where these metrics are used to guide code regeneration. ("Iterative Code Improvement" in Figure \ref{fig:fig1})
\subsubsection{RQ4 Design} In this research question, we extend our complexity-aware feedback method to an agent-based code generation framework. Specifically, we first apply Reflexion, a feedback-driven code generation agent, to iteratively refine the generated code. Once Reflexion has completed its improvement process, we then incorporate our complexity-based feedback mechanism to further enhance the correctness of the code.

Our approach remains consistent with RQ3, where we identify the most predictive complexity metrics using Shapley values and iteratively prompt the LLM to adjust its generated code based on these insights. However, in RQ4, this complexity-driven refinement is applied on top of Reflexion’s intermediate outputs rather than directly on the initial code generated by the LLM. This allows us to evaluate whether our complexity-aware feedback can further improve results even when an agent has already optimized the code.

Given that datasets like HumanEval, MBPP, and LeetCode already exhibit high Pass@1 scores, we focus on BigCodeBench, a more challenging dataset where there is greater room for improvement. By layering our complexity-aware approach on top of an agent-based method, we assess its effectiveness in refining code generation within more complex and practical programming scenarios.

\begin{algorithm}[ht!]
\caption{Complexity-Aware Feedback for Enhanced LLM Code Generation}
\label{alg:alg1}
\begin{algorithmic}[1]
\REQUIRE Training set $T$, Evaluation set $E$, Max iterations $I = 5$
\ENSURE Final pass@1 scores

\FOR{each code $c$ in $T$}
    \STATE $g \gets \text{GenerateCode(LLM, } c \text{)}$
    \STATE $\text{metrics}_g \gets \text{ComputeComplexityMetrics(} g \text{)}$
    \STATE $\text{pass1}_g \gets \text{CalculatePass1(} g \text{)}$
\ENDFOR
\STATE $\text{important\_metrics} \gets$ \text{FindImportantMetrics(}
\STATE \quad \text{LogisticRegression(} $T$ \text{), ShapleyValues)}

\FOR{each code $c$ in $E$}
    \STATE $g \gets \text{GenerateCode(LLM, } c \text{)}$
    \STATE $\text{pass1}_g \gets \text{EvaluatePass1(} g, \text{GeneratedTestCases)}$

    \IF{$\text{pass1}_g == 0$}
        \STATE $\text{metrics}_g \gets \text{ComputeComplexityMetrics(} g \text{)}$
        \FOR{$i = 1$ to $I$}
            \STATE \textbf{Prompt LLM:} \textit{Please ensure that your generated code has different values for the following complexity metrics:} $\mathcal{M}$
            \STATE $g \gets \text{RegenerateCode(LLM, }$
            \STATE \quad \text{metrics = important\_metrics)}
            \STATE $\text{pass1}_g \gets \text{EvaluatePass1(} g, \text{GeneratedTestCases)}$
            \IF{$\text{pass1}_g == 1$}
                \STATE \textbf{break}
            \ENDIF
        \ENDFOR
    \ENDIF
\ENDFOR

\FOR{each code $g$ in $E$}
    \STATE $\text{final\_pass1} \gets \text{EvaluatePass1(} g, \text{OriginalTestCases)}$
\ENDFOR

\end{algorithmic}
\end{algorithm}

%% file: sec/6_Results.tex
\section{Experiment Results}

\subsection{RQ1: Are complexity metrics of the generated codes correlated with the code generation's effectiveness (pass@1)?}


Table \ref{tab:LR_results} presents the Logistic Regression model's accuracy (the average accuracy over five folds) when using different feature selection methods for each LLM on the four datasets. The bold values in each row represent the highest accuracy achieved for that particular LLM and dataset combination. For example, the highest accuracy for GPT-4o on the HumanEval dataset is 0.921 using Shapley Values, indicating that Shapley Values outperformed other methods for this combination.

\begin{table*}[t!]
\centering
\caption{The accuracy of the logistic regression model when using different
feature selection methods for each LLM on the four datasets. ``-'' indicates no feature selection.}
\label{tab:LR_results}
\begin{tabular}{|l|l|l|l|l|l|}
\hline \textbf{Dataset + Model/ feature selection method} & - & L1 Regularization & RFE & Correlation & Shapley\\
\hline HumanEval + GPT4o & 0.872 & 0.915 & 0.903 & 0.909 & $\mathbf{0 . 9 2 1}$ \\
\hline HumanEval + GPT-3.5-turbo & 0.566 & 0.589 & $\mathbf{0 . 6 8 3}$ & 0.662 & 0.678 \\
\hline HumanEval + Llama 3.1 & 0.651 & $\mathbf{0 . 7 1 4}$ & 0.665 & 0.708 & 0.689 \\
\hline
\hline MBPP + GPT4o & 0.680 & 0.701 & 0.712 & 0.708 & $\mathbf{0 . 7 4}$ \\
\hline MBPP+ GPT-3.5-turbo & 0.607 & 0.607 & $\mathbf{0 . 6 4 5}$ & 0.638 & 0.623 \\
\hline MBPP+ Llama 3.1 & 0.635 & 0.635 & 0.622 & 0.654 & $\mathbf{0 . 6 6 2}$\\
\hline
\hline LeetCode + GPT4o & 0.895 & 0.895 & 0.893 & 0.895 & $\mathbf{0 . 8 9 7}$ \\
\hline LeetCode + GPT-3.5-turbo & 0.805 & $\mathbf{0 . 8 1 4}$ & 0.805 & 0.791 & 0.811 \\
\hline LeetCode + Llama 3.1 & 0.642 & 0.668 & 0.661 & $\mathbf{0 . 6 9 3}$ & 0.686 \\
\hline
\hline BigCodeBench + GPT-o3-mini & 0.432 & 0.562 & 0.560 & $\mathbf{0 . 5 9 5}$ & 0.587 \\
\hline BigCodeBench + GPT4o & 0.648 & 0.681 & 0.702 & 0.695 & $\mathbf{0 . 7 2 2}$ \\
\hline
\end{tabular}

\end{table*}

Table \ref{tab:LR_results} reveals clear patterns in the relationship between the complexity metrics of LLM-generated code and pass@1 scores across different datasets and LLMs. The key insights include:

\noindent \textbf{HumanEval Dataset:} This dataset consistently showed the best results across all LLMs, particularly with GPT-4o, which achieved the highest accuracy (0.921). This indicates a strong correlation between the complexity metrics and pass@1 for this dataset, especially with more advanced models like GPT-4o. 
Even less sophisticated models, such as GPT-3.5-turbo and Llama 3.1 showed significant improvement on HumanEval, achieving accuracies of 0.683 and 0.714, respectively. This suggests that the complexity metrics are particularly well-suited to predicting pass@1 for the HumanEval dataset across different LLMs.

\noindent \textbf{MBPP Dataset:} 
Performance on MBPP was weak across all LLMs, with GPT-3.5 Turbo and Llama 3.1 struggling at an accuracy below 0.67. Even GPT-4o, which excelled elsewhere, achieved only 0.74, possibly due to the nature of the tasks in MBPP, which may involve more challenging or less uniform code patterns, making complexity metrics less predictive of Pass@1.


\noindent \textbf{LeetCode Dataset} 
LeetCode presented an interesting middle ground, with GPT-4o performing consistently well across all methods, demonstrating that the complexity metrics are robust for this dataset.
GPT-3.5-turbo and Llama 3.1 also performed moderately well, with GPT-3.5-turbo reaching 0.814 and Llama 3.1 achieving 0.693. These results suggest that while LeetCode is not as easy to predict as HumanEval, it still allows for a solid correlation between complexity metrics and pass@1.

\noindent \textbf{BigCodeBench Dataset} 
BigCodeBench exhibited the weakest overall performance, with GPT-o3 mini achieving only 0.59 accuracy and GPT-4o reaching 0.722, lower than its performance on other datasets. These results suggest that complexity metrics are less predictive of Pass@1 in this dataset, likely due to its diverse function calls and more intricate coding tasks. This highlights the need for alternative or enhanced feedback mechanisms for handling more complex real-world programming challenges.

Compared to other feature selection methods, Shapley values perform best, as shown by the higher number of top-performing values in the Shapley column. Even where Shaply is not the top method, its accuracy is very close to the best-performing alternatives. This consistency is why we chose Shaply for feature selection in our feedback-based algorithm.


\mybox{\textbf{Answer to RQ1:} The results indicate that complexity metrics of LLM-generated code solutions are correlated with their pass@1 scores, particularly when feature selection methods such as Shapley Values are applied.} 

\subsection{RQ2: How do the distributions of complexity metrics differ between successful and failed solutions generated by LLMs?}

\begin{figure}
    \centering
    \includegraphics[width=0.5\linewidth]{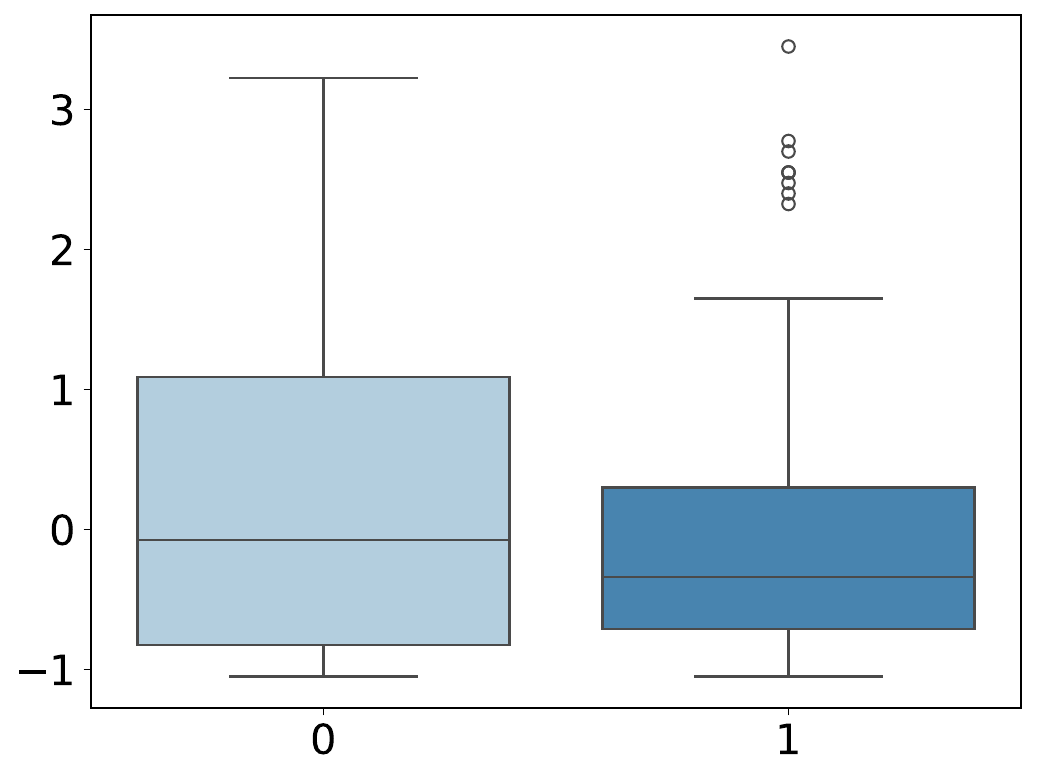}
    \caption{The distribution of Halstead Length by target value (pass@1 = 1 or pass@1 = 0), using GPT-4o as the LLM and HumanEval as the dataset.}
    \label{fig:fig2}
\end{figure}

In this research question, we aim to explore the differences in the distributions of complexity metrics based on the target value (pass@1 = 1 or pass@1 = 0). For example, Figure \ref{fig:fig2} presents a box plot of the Halstead Length distribution, comparing the target values when GPT-4o is used as the LLM and HumanEval as the dataset. The median for target 0 is higher than for target 1, indicating that this complexity metric tends to be greater in cases where the code fails.

\begin{figure*}[ht!]
   \subfloat[Dataset: HumanEval, LLM: gpt4o \label{fig:fig3.1}]{%
      \includegraphics[ width=0.3\textwidth]{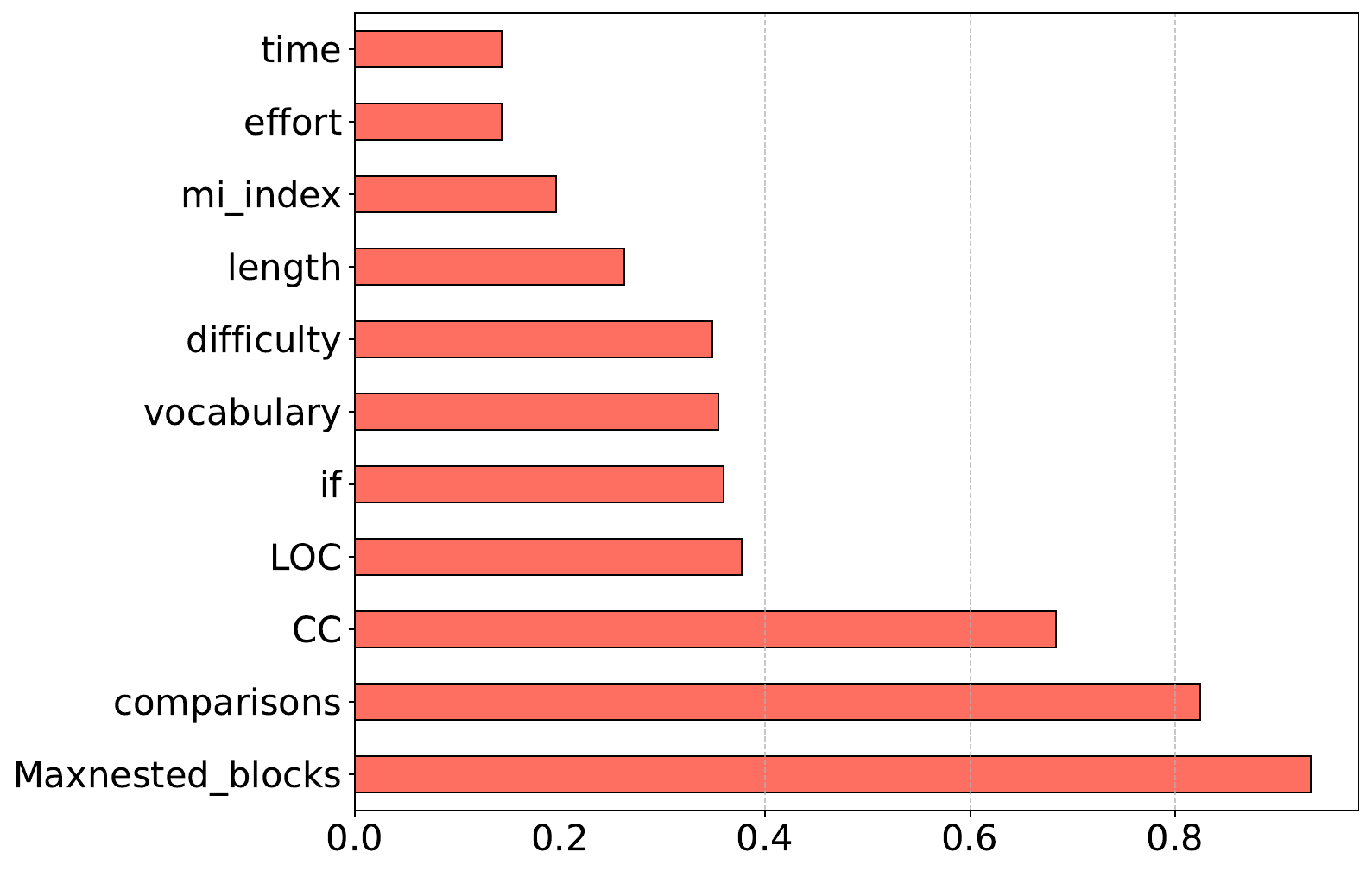}}
\hspace{\fill}
   \subfloat[Dataset: HumanEval, LLM: gpt3.5  \label{fig:fig3.2} ]{%
      \includegraphics[ width=0.3\textwidth]{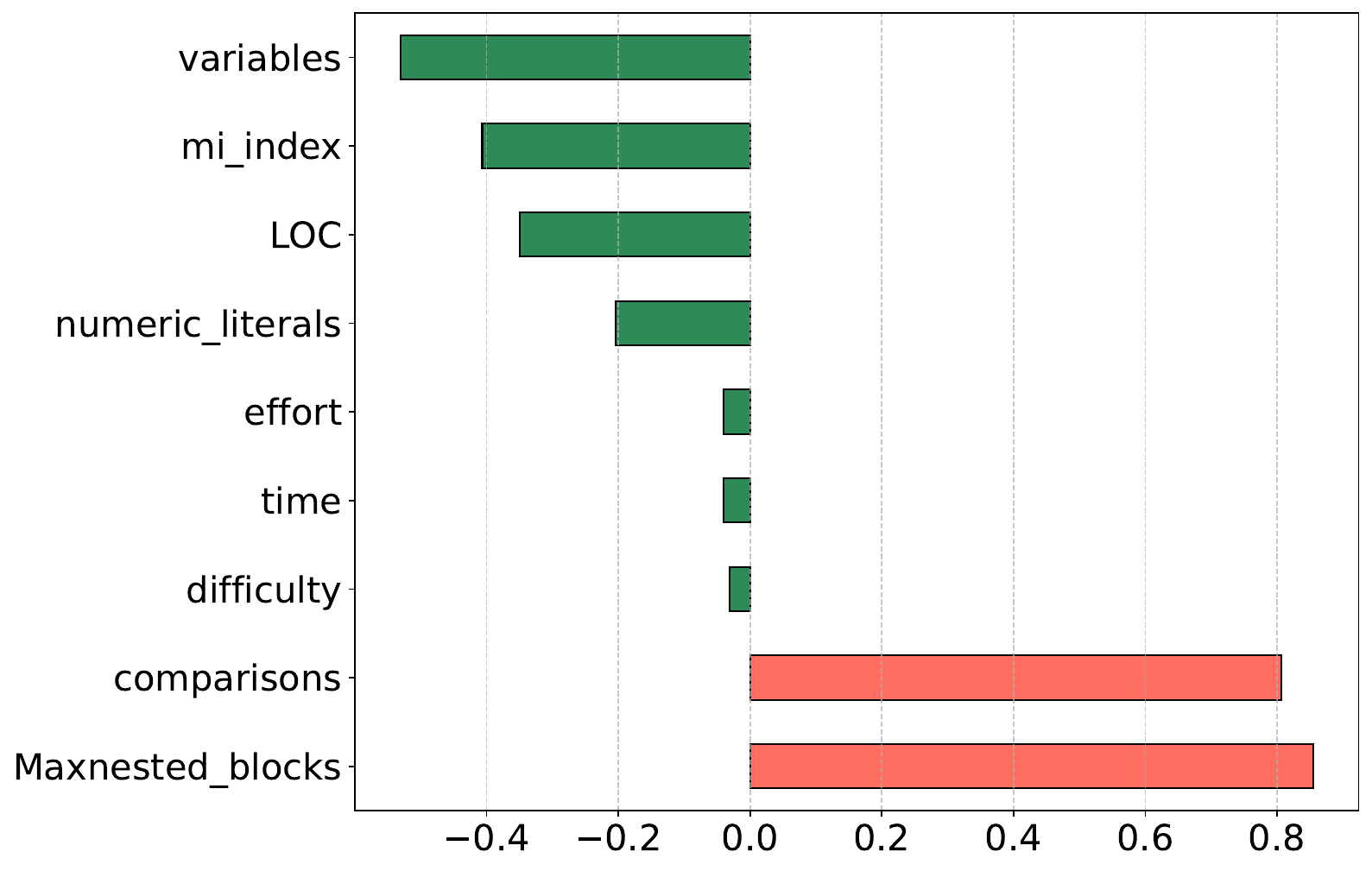}}
\hspace{\fill}
   \subfloat[Dataset: HumanEval, LLM: Llama3.1 \label{fig:fig3.3}]{%
      \includegraphics[ width=0.3\textwidth]{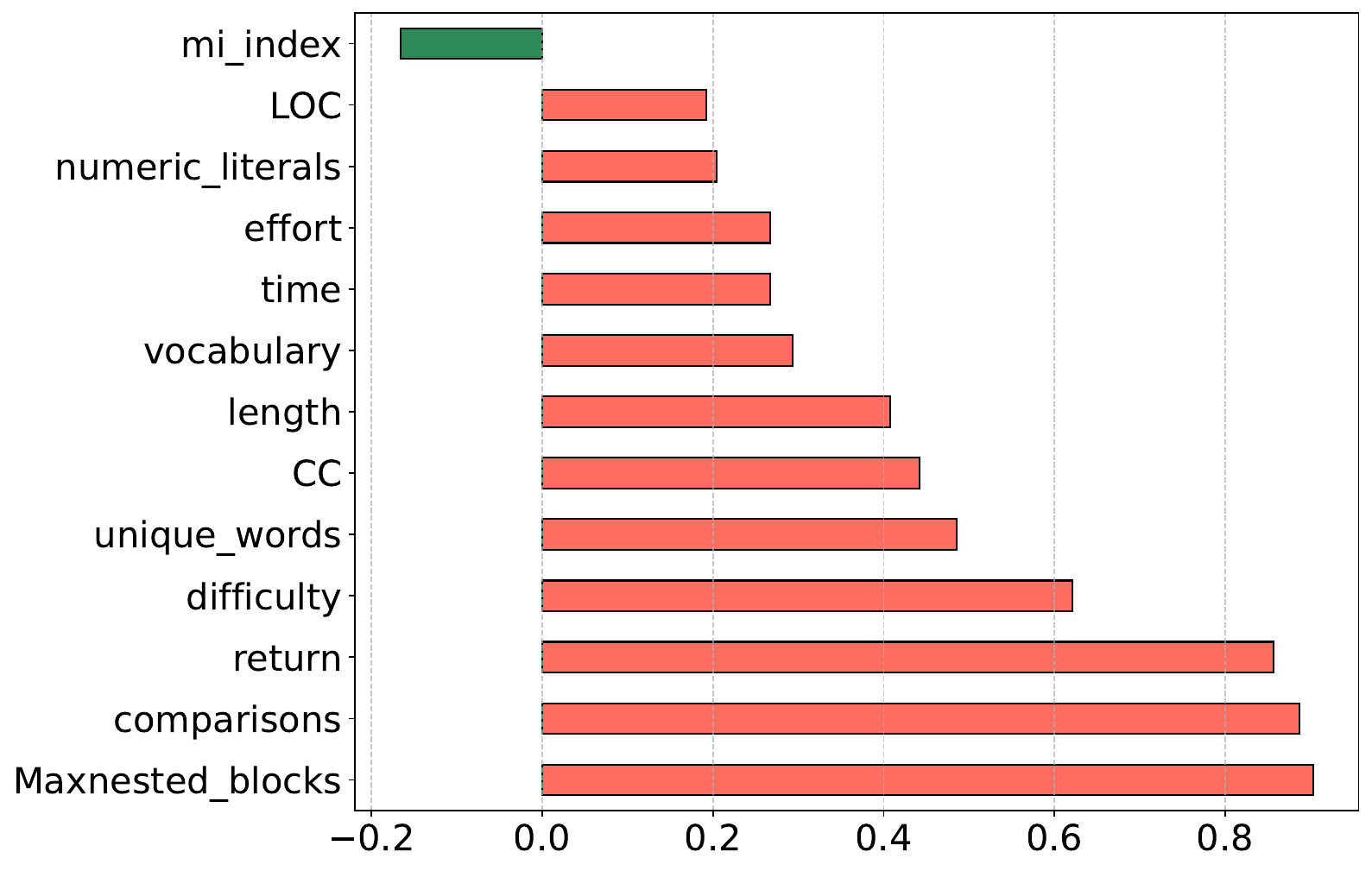}}\\

       \subfloat[Dataset: MBPP-sanitized, LLM: gpt4o \label{fig:fig3.4}]{%
      \includegraphics[ width=0.3\textwidth]{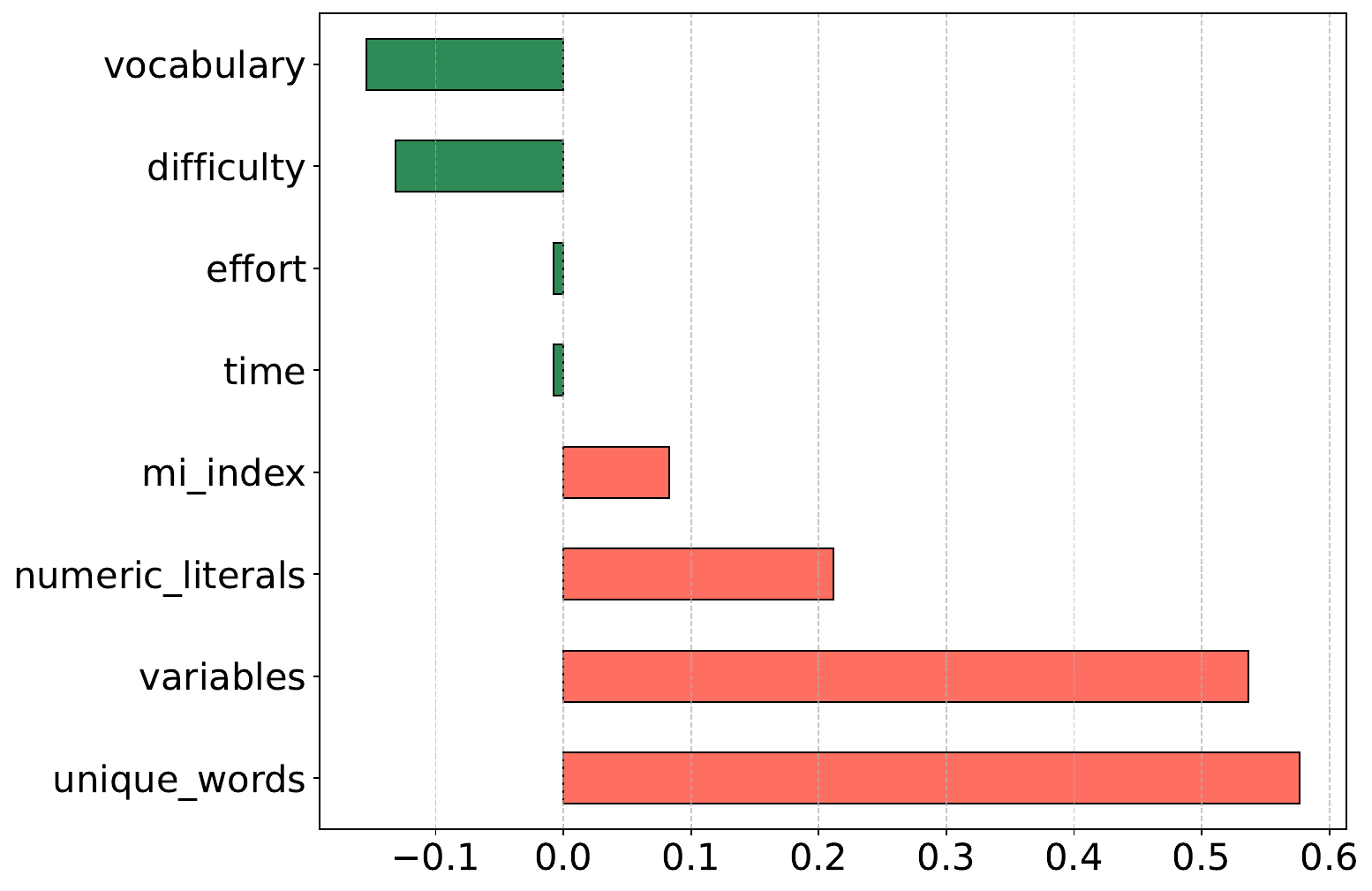}}
\hspace{\fill}
   \subfloat[Dataset: MBPP-sanitized, LLM: gpt3.5  \label{fig:fig3.5} ]{%
      \includegraphics[ width=0.3\textwidth]{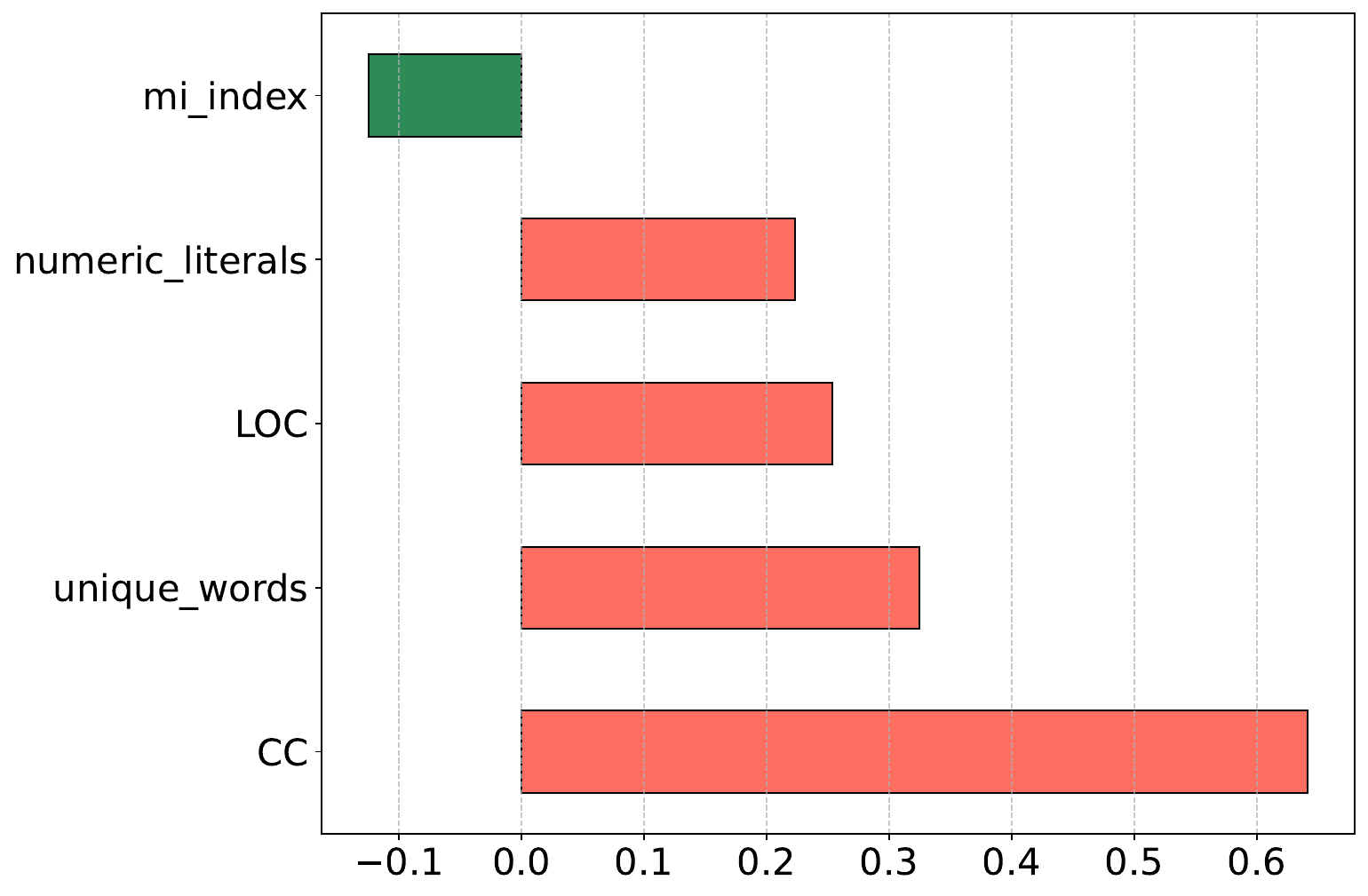}}
\hspace{\fill}
   \subfloat[Dataset: MBPP-sanitized, LLM: Llama3.1 \label{fig:fig3.6}]{%
      \includegraphics[ width=0.3\textwidth]{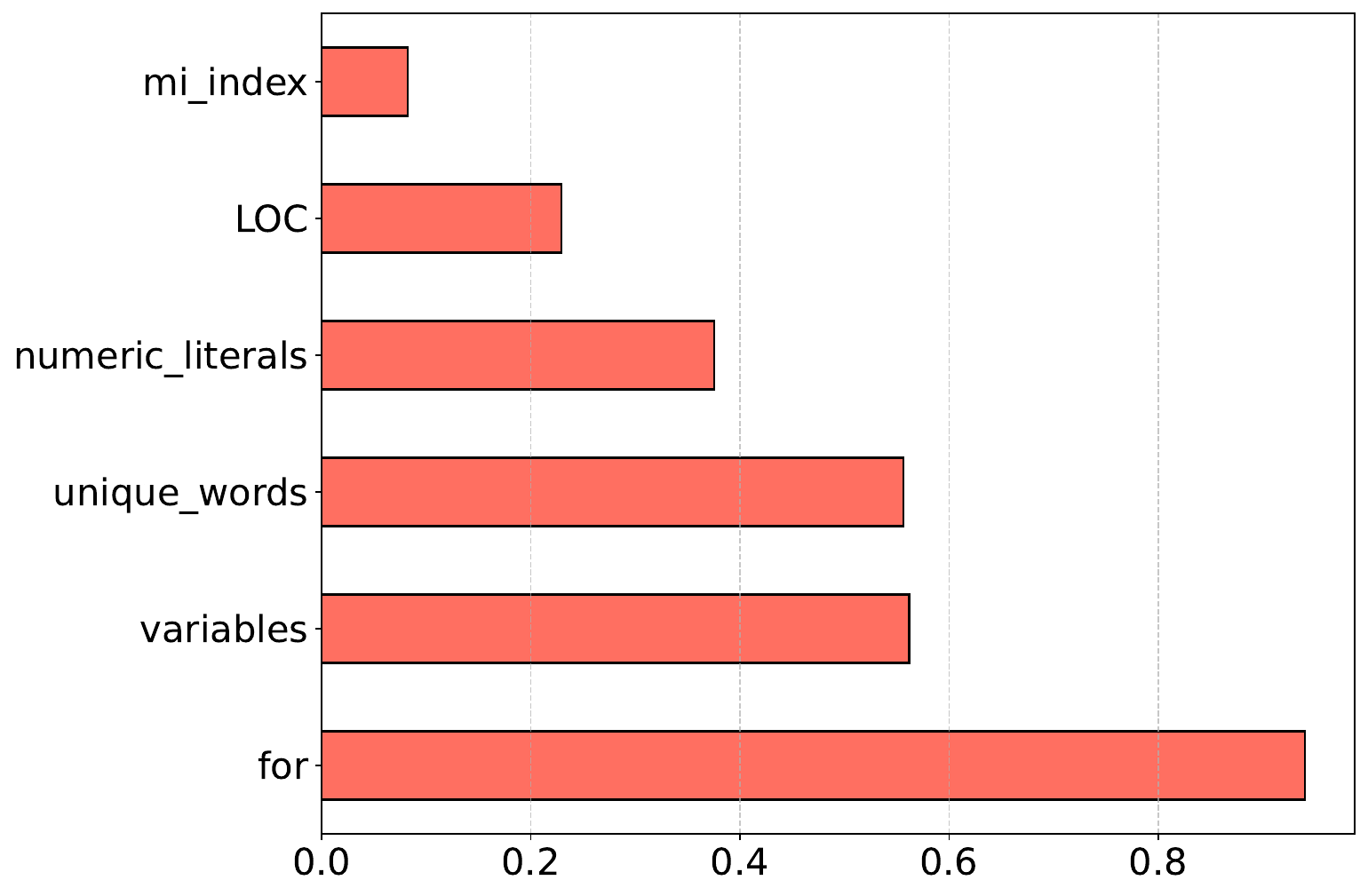}}\\

       \subfloat[Dataset: LeetCode, LLM: gpt4o \label{fig:fig3.7}]{%
      \includegraphics[ width=0.3\textwidth]{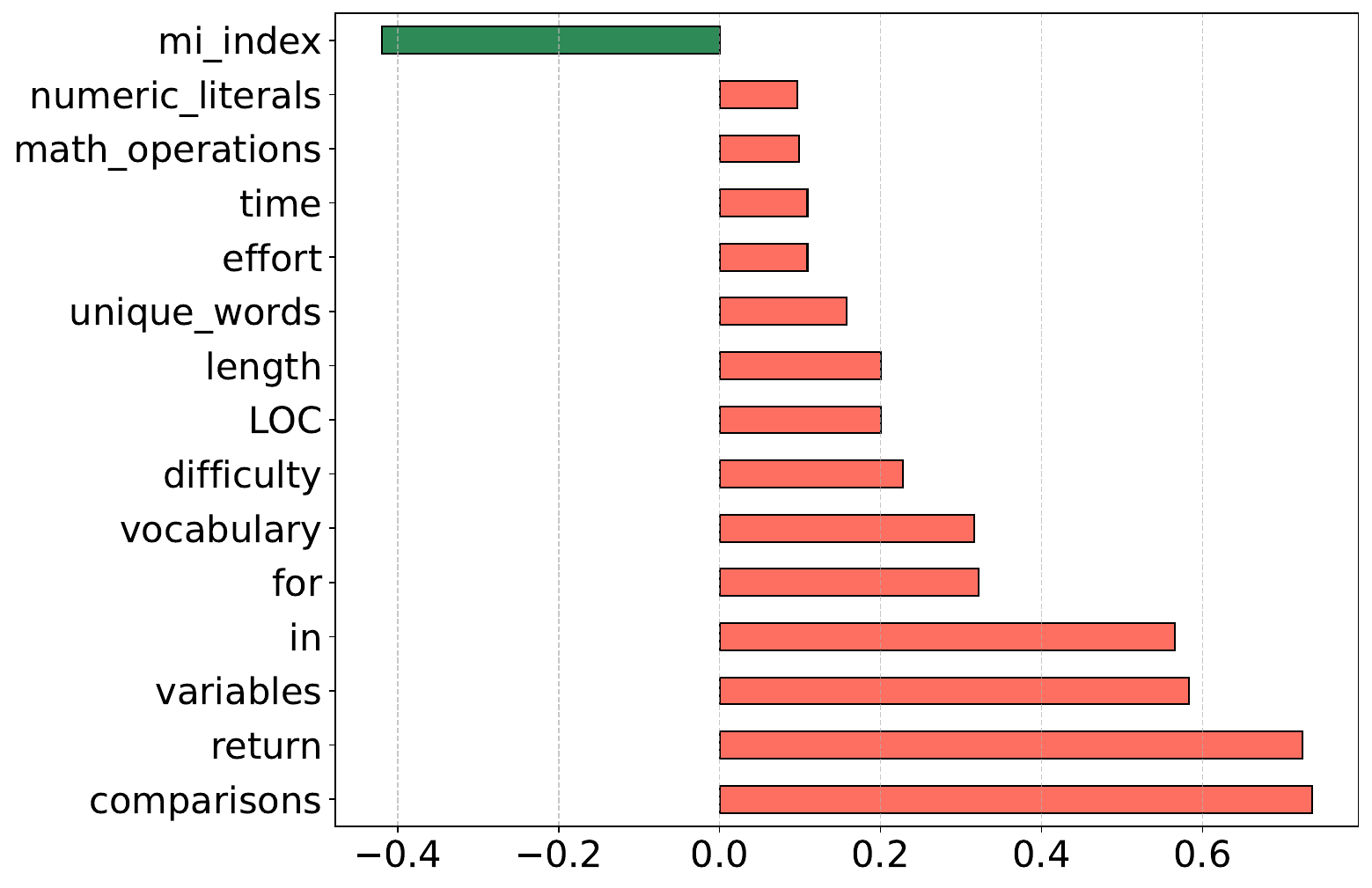}}
\hspace{\fill}
   \subfloat[Dataset: LeetCode, LLM: gpt3.5  \label{fig:fig3.8} ]{%
      \includegraphics[ width=0.3\textwidth]{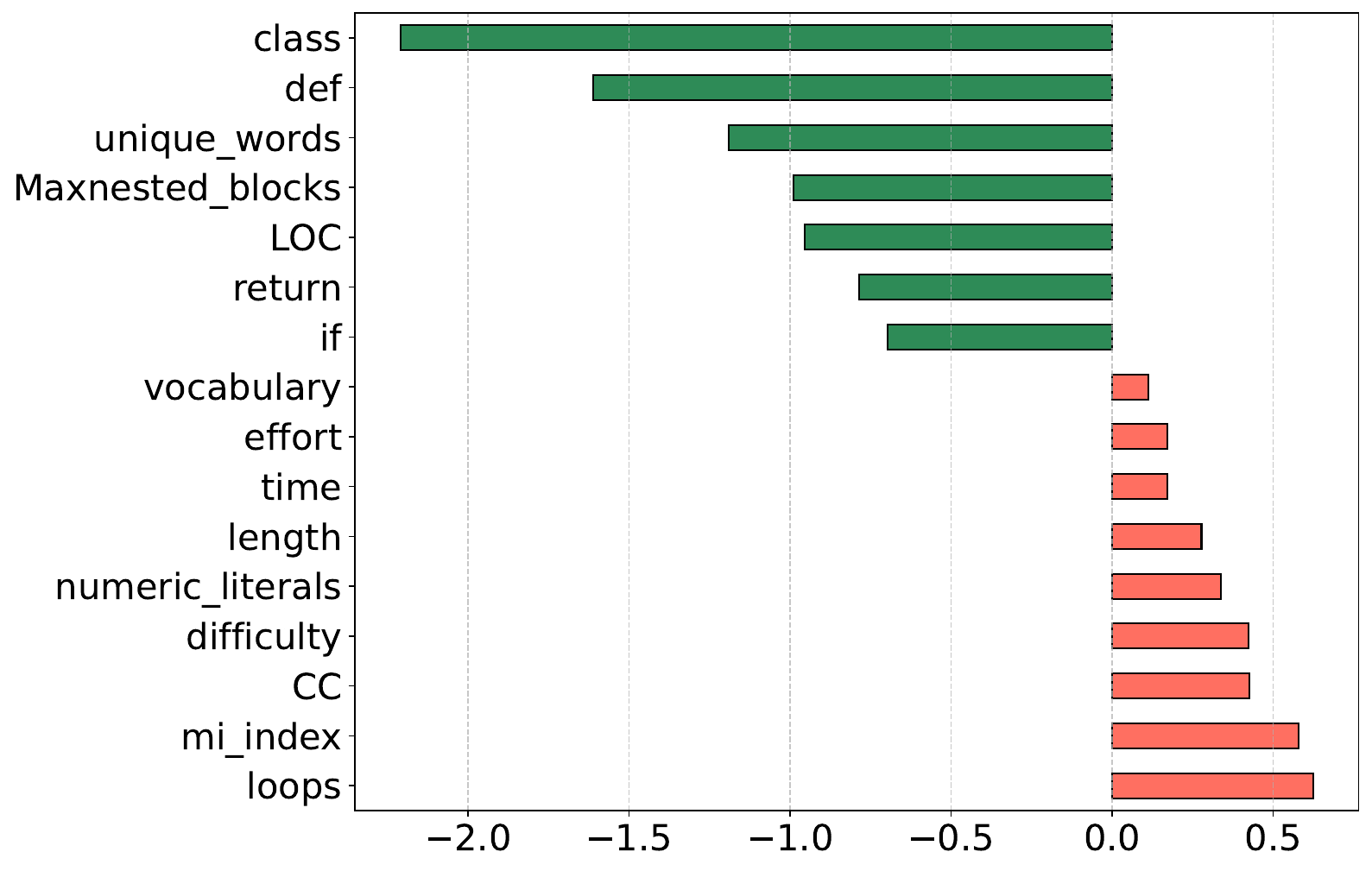}}
\hspace{\fill}
   \subfloat[Dataset: LeetCode, LLM: Llama3.1 \label{fig:fig3.9}]{%
      \includegraphics[ width=0.3\textwidth]{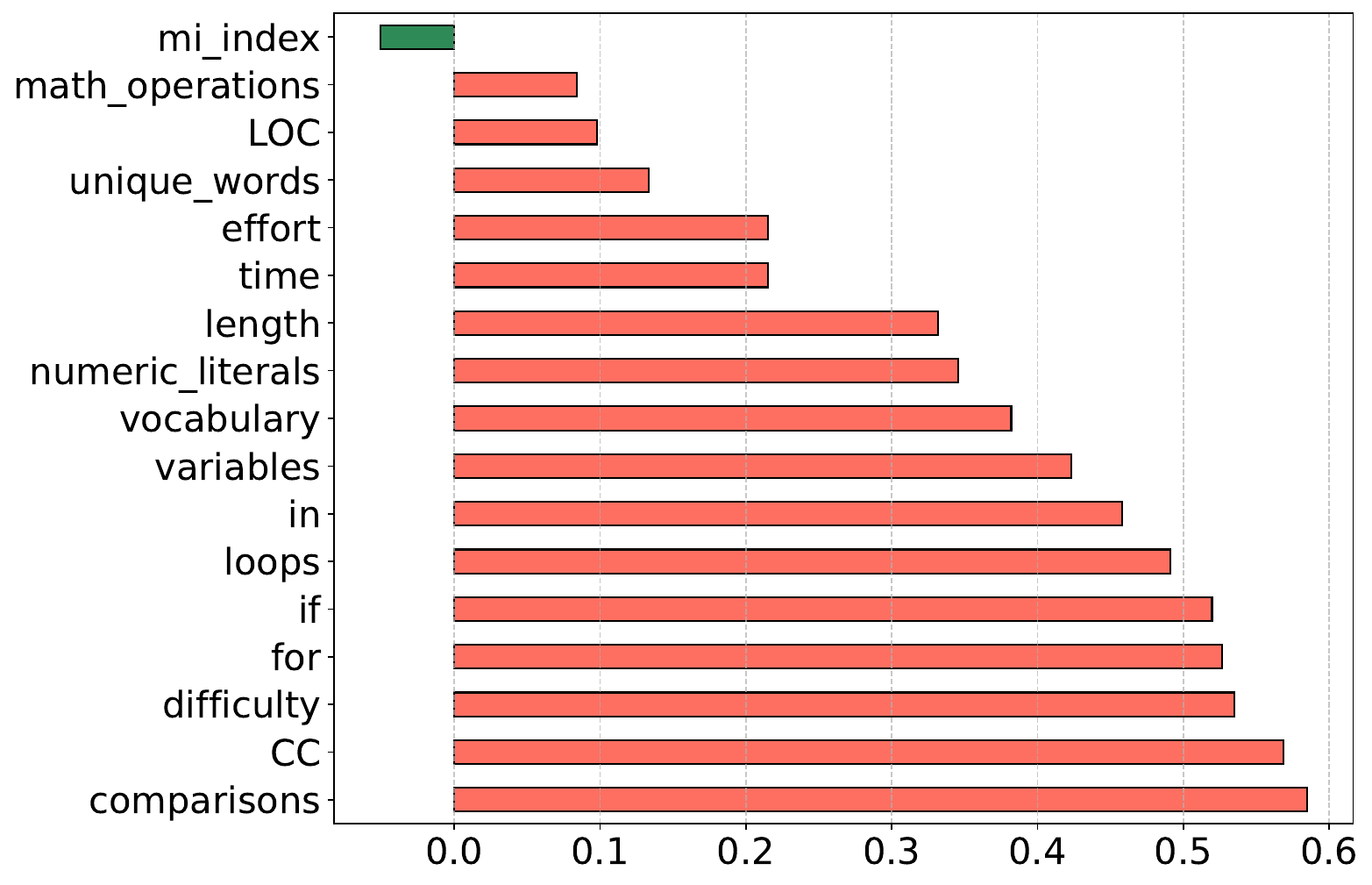}}\\

\caption{Bar plot showing the median differences in complexity metrics between target values (pass@1 = 1 and pass@1 = 0) for each LLM and dataset combination. (the median difference in metrics that are not in the plots is zero)}
    \label{fig:fig3}
\end{figure*}

To highlight the variation between the two target distributions, we calculate the difference in their medians for each complexity metric and visualize these differences using a bar plot. 
Figure~\ref{fig:fig3} displays this bar plot for each LLM and dataset combination. 


\subsubsection{RQ2.1: Are there specific metrics that LLMs have more difficulty getting right when generating code?}

In Figure \ref{fig:fig3}, each column presents the difference in the median values of complexity metrics between failed and passed cases across different datasets, evaluated using a specific LLM. A clear pattern emerges: GPT-4o and Llama 3.1 consistently generate more complex code in failed cases (pass@1 = 0), with metrics such as Halstead Length and number of lines in the code (LOC) showing higher values when code fails. This suggests that these two models might produce overly complex solutions, which could lead to incorrect code outputs.

In contrast, GPT-3.5-turbo tends to generate simpler code for failed cases, as indicated by several complexity metrics being higher for the correct cases (pass@1 = 1). This may indicate that GPT-3.5-turbo generates overly simplistic solutions that lack the required details to pass.

\subsubsection{RQ2.2: Do different datasets' characteristics differ in terms of code complexity metrics' distributions over the correct vs. incorrect code?}

Each row in Figure \ref{fig:fig3} represents the behavior of a dataset across multiple LLMs. A notable finding is that both the HumanEval and LeetCode datasets show more variation in complexity metrics between failed and successful cases. This suggests that for these datasets, the complexity of the code plays a significant role in determining whether a generated solution passes or fails. 

On the other hand, the MBPP dataset shows fewer differences in complexity metrics between pass@1 = 1 and pass@1 = 0. This finding is consistent with the lower performance seen in RQ1 for this dataset, where the accuracy of the Logistic Regression model was also lower. It appears that complexity metrics are less relevant to determining code correctness in MBPP, which aligns with the nature of MBPP prompts—these are typically one-line explanations without detailed context, potentially causing issues beyond the complexity of the generated code itself.  
In contrast, HumanEval and LeetCode provide more comprehensive prompts, including examples and details, meaning that code failures are more likely related to the LLM’s handling of code complexity. In such cases, offering complexity-based feedback could lead to performance improvements, as indicated by the differences in metric distributions. 


\textbf{HumanEval Dataset:} Across all three LLMs, failed code tends to have higher maximum nested blocks and more comparisons. This suggests that in HumanEval, incorrect code is often more structurally complex, containing deeper levels of nesting and a greater number of comparison operations. 
Therefore, excessive nesting and comparisons add complexity and decision points, increasing the likelihood of errors and indicating potential code failure.

\textbf{MBPP Dataset:} For all three LLMs, failed code exhibits a higher number of unique words and more numeric literals. 
This indicates that incorrect MBPP solutions introduce more distinct terms and numeric values. Since MBPP prompts are typically concise, failed solutions likely overcomplicate the task by adding extra variables or values that do not align with the intended simplicity. This increased complexity may contribute to the failure of the generated code.

\textbf{LeetCode Dataset:} In this dataset, failed code 
has higher counts of numeric literals and consistently higher values across all Halstead complexity metrics. This pattern suggests that incorrect solutions in LeetCode tend to be more computationally demanding and complex. The prominence of Halstead metrics, which quantify the overall effort required to understand and maintain code, suggests that these failed solutions are not only larger but also harder to process and may involve more intricate logic or operations than necessary, which contributes to their failure.

\mybox{\textbf{Answer to RQ2:} 
The analysis reveals that complexity patterns differ across models and datasets: GPT-4o and Llama 3.1 struggle with complex outputs, while GPT-3.5-turbo tends to fail on overly simple ones. Complexity strongly impacts failures in HumanEval and LeetCode but is less relevant in MBPP due to its simpler prompts.}

\subsection{RQ3: Can feedback based on complexity metric values of the generated code improve LLMs' code generation effectiveness?}

\begin{figure}
    \centering
    \includegraphics[width=0.9\linewidth]{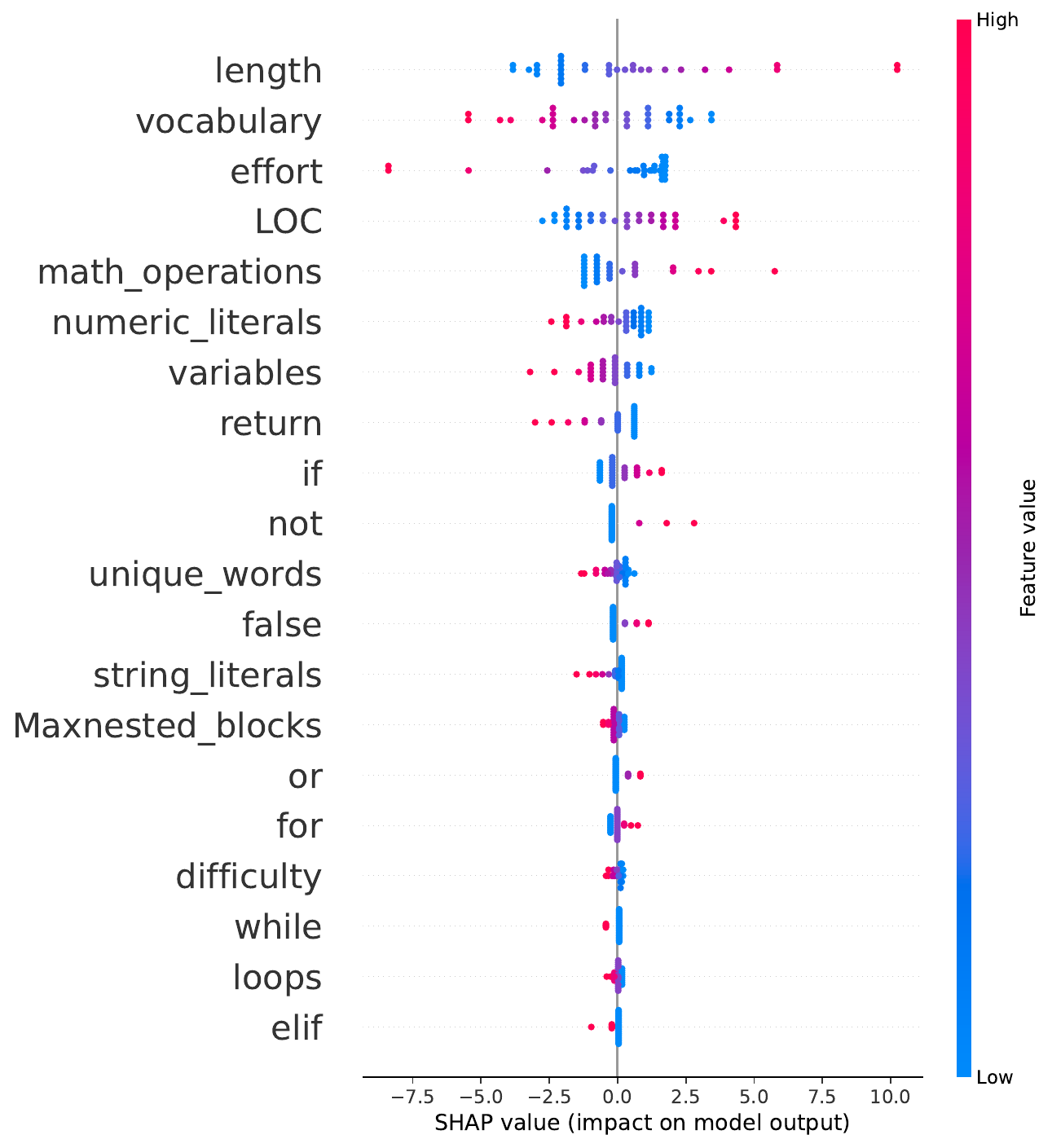}
    \caption{Shapley values illustrating the importance of various complexity metrics in predicting the likelihood of generated code passing all test cases (pass@1) for one fold of the HumanEval dataset, using GPT-4o for code generation. The chart highlights the most influential metrics.} 
    \label{fig:fig4}
\end{figure}

In this research question, we focus on improving code generation by providing targeted feedback to LLMs based on key complexity metrics. Table \ref{tab:tab2} shows the Pass@1 performance across iterations for each LLM and dataset. Iteration 0 serves as a baseline where the LLM generates code without feedback. 
From Iteration 1, the LLM is prompted to modify the five most important complexity metrics for any incorrect code solution, identified during training with logistic regression and Shapley values. This process continues for five iterations, refining the generated code. 
For example, Figure \ref{fig:fig4} shows the most influential metrics using Shapley values calculated from one fold of the HumanEval dataset, with code generated by GPT-4o. The metrics displayed are ranked according to their impact on the model's predictions. This ranking facilitates easy identification of which complexity features contribute most significantly to the likelihood of success. Metrics such as Halstead Length, vocabulary, effort, Lines of Code (LOC), and number of math operations emerged as the most significant indicators of code quality. For the MBPP dataset, we used the provided training set during the training phase. However, for HumanEval and LeetCode, which do not include separate training sets, we employed 5-fold cross-validation and reported the average results across folds. 

The second baseline in Table \ref{tab:tab2}, represented by the white rows, involves asking the LLM to regenerate incorrect codes iteratively without any complexity-based feedback. This setup allows us to assess whether our approach offers improvements over naive regeneration. In this baseline, we follow the same process as our method, using LLM-generated test cases for evaluation during the iterations. 

It is important to note that while our algorithm incorporates LLM-generated test cases for feedback during training, we report the final Pass@1 results based on the actual test cases from the dataset.

\begin{table*}[t!]
    \centering
        \caption{Pass@1 of our feedback-based approach (green rows) and the baseline (white rows) for various LLMs on HumanEval, MBPP, and LeetCode datasets}
    \label{tab:tab2}
    \arrayrulecolor{black}
    \setlength{\arrayrulewidth}{0.5mm}
    \renewcommand{\arraystretch}{1.3}  
    \begin{adjustbox}{max width=\textwidth}
    \begin{tabular}{|l|l|l|c|c|c|c|c|c|}
        \hline
        \textbf{Model} & \textbf{Dataset} & \textbf{Complexity Metrics} & \textbf{Pass@1 (Iter 0)} & \textbf{Pass@1 (Iter 1)} & \textbf{Pass@1 (Iter 2)} & \textbf{Pass@1 (Iter 3)} & \textbf{Pass@1 (Iter 4)} & \textbf{Pass@1 (Iter 5)} \\
        \hline
        \rowcolor{green!20} gpt4o & HumanEval (5-fold) & 5 most effective & 0.89 & 0.93 & 0.93 & 0.95 & 0.94 & 0.95 \\
        gpt4o & HumanEval (5-fold) & - & 0.89 & 0.92 & 0.92 & 0.91 & 0.91 & 0.91 \\
        \rowcolor{green!20} gpt3.5-turbo & HumanEval (5-fold) & 5 most effective & 0.56 & 0.72 & 0.75 & 0.75 & 0.75 & 0.76 \\
        gpt3.5-turbo & HumanEval (5-fold) & - & 0.56 & 0.60 & 0.61 & 0.63 & 0.62 & 0.63 \\
        \rowcolor{green!20} Llama 3.1 & HumanEval (5-fold) & 5 most effective & 0.68 & 0.73 & 0.73 & 0.74 & 0.75 & 0.75 \\
        Llama 3.1 & HumanEval (5-fold) & - & 0.68 & 0.70 & 0.71 & 0.71 & 0.71 & 0.71 \\
        \hline
        \rowcolor{green!20} gpt4o & MBPP-sanitized & 5 most effective & 0.68 & 0.71 & 0.71 & 0.72 & 0.72 & 0.72 \\
        gpt4o & MBPP-sanitized & - & 0.70 & 0.70 & 0.71 & 0.71 & 0.72 & 0.71 \\
        \rowcolor{green!20} gpt3.5-turbo & MBPP-sanitized & 5 most effective & 0.67 & 0.69 & 0.70 & 0.70 & 0.70 & 0.70 \\
        gpt3.5-turbo & MBPP-sanitized & - & 0.67 & 0.66 & 0.68 & 0.67 & 0.68 & 0.68 \\
        \rowcolor{green!20} Llama 3.1 & MBPP-sanitized & 5 most effective & 0.55 & 0.56 & 0.57 & 0.57 & 0.59 & 0.60 \\
        Llama 3.1 & MBPP-sanitized & - & 0.55 & 0.56 & 0.57 & 0.57 & 0.59 & 0.59 \\
        \hline
        \rowcolor{green!20} gpt4o & LeetCode (5-fold) & 5 most effective & 0.91 & 0.91 & 0.92 & 0.92 & 0.92 & 0.92 \\
        gpt4o & LeetCode (5-fold) & - & 0.92 & 0.91 & 0.92 & 0.92 & 0.92 & 0.92 \\
        \rowcolor{green!20} gpt3.5-turbo & LeetCode (5-fold) & 5 most effective & 0.81 & 0.83 & 0.84 & 0.84 & 0.85 & 0.85 \\
        gpt3.5-turbo & LeetCode (5-fold) & - & 0.81 & 0.81 & 0.82 & 0.82 & 0.83 & 0.82 \\
        \rowcolor{green!20} Llama 3.1 & LeetCode (5-fold) & 5 most effective & 0.66 & 0.67 & 0.69 & 0.70 & 0.70 & 0.71 \\
        Llama 3.1 & LeetCode (5-fold) & - & 0.66 & 0.67 & 0.68 & 0.67 & 0.68 & 0.68 \\
        \hline
    \end{tabular}
    \end{adjustbox}
\end{table*}

The most substantial improvements are observed in the HumanEval dataset, particularly with GPT-3.5, which initially had the lowest pass@1 but achieved a higher pass@1 than Llama 3 in the fifth iteration. This result aligns with our findings in RQ1 and RQ2, confirming that pass@1 is correlated with the complexity metrics of generated code. Additionally, our algorithm outperformed the baseline across all models. Specifically, GPT-4o’s pass@1 improved by 6.74\% with our algorithm versus 2.24\% in the baseline, GPT-3.5-turbo by 35.71\% versus 12.5\%, and Llama 3.1 by 10.29\% versus 4.41\%.

In the MBPP dataset, while our algorithm achieved better results than the zero-shot baseline, the improvements were only slightly higher than in the second baseline, where complexity metrics were not directly leveraged. This outcome aligns with previous RQs, which indicated that MBPP exhibits a low correlation between complexity metrics and pass@1, making complexity-based adjustments less impactful. Improvements for GPT-4o were 5.88\% with our algorithm (versus 1.43\% in the baseline), for GPT-3.5-turbo 4.48\% (versus 1.49\%), and for Llama 3.1 9.09\% (versus 7.27\%).

In the LeetCode dataset, GPT-4o showed limited improvement, likely due to its initially high pass@1 score, suggesting it may have reached its performance ceiling in this context. For instance, this is a prompt in the LeetCode dataset: \textit{``Given a string s, return \_all the palindromic permutations (without duplicates) of it\_. You may return the answer in **any order**. If `s` has no palindromic permutation, return an empty list.''} Despite mentioning ``any order'', order errors persisted, implying that these failures were more due to prompt interpretation issues than complexity. Conversely, improvements on GPT-3.5-turbo and Llama 3.1 were more pronounced. The improvement for GPT-4o was 1.09\% (versus no improvement in the baseline), for GPT-3.5-turbo, was 4.94\% (versus 1.23\%), and for Llama 3.1, was 7.58\% (versus 3.03\%).

Across all datasets, GPT-3.5 consistently showed the greatest improvement relative to other LLMs. This could be because when we ask the LLM to change the complexity metrics, it will usually make the code more complex than more simple because our observations in RQ2 suggest that GPT-3.5 initially generates simpler code in failed cases, whereas GPT-4o and Llama 3.1 produce more complex code upon failure.

\mybox{\textbf{Answer to RQ3:} Complexity-based feedback significantly improves LLMs' code generation ability, particularly for GPT-3.5-turbo, which saw notable gains in Pass@1 compared to baselines. By refining key complexity metrics, our approach consistently outperformed the baseline, confirming that targeted complexity-based adjustments can enhance the accuracy of generated codes.}
\subsection{RQ4: Can feedback based on complexity metric values enhance the effectiveness of code generation agents?}

\begin{table*}[t!]
    \centering
        \caption{Pass@1 of our complexity-based feedback approach (green rows) and the baseline (white rows) for GPT-4o and GPT-o3 mini on the BigCodeBench dataset, with and without the Reflexion agent}
    \label{tab:tab3}
    \arrayrulecolor{black}
    \setlength{\arrayrulewidth}{0.5mm}
    \renewcommand{\arraystretch}{1.3}  
    \begin{adjustbox}{max width=\textwidth}
    \begin{tabular}{|l|l|l|l|c|c|c|c|c|c|}
        \hline
        \textbf{Model} & \textbf{Dataset} & \textbf{Agent} & \textbf{Complexity Metrics} & \textbf{Pass@1 (Iter 0)} & \textbf{Pass@1 (Iter 1)} & \textbf{Pass@1 (Iter 2)} & \textbf{Pass@1 (Iter 3)} & \textbf{Pass@1 (Iter 4)} & \textbf{Pass@1 (Iter 5)} \\
        \hline
        \rowcolor{green!20} gpt4o & BigCodeBench (5-fold) & - & 5 most effective & 0.28 & 0.30 & 0.32 & 0.33 & 0.34 & 0.34 \\
        gpt4o & BigCodeBench (5-fold) & - & - & 0.28 & 0.28 & 0.30 & 0.30 & 0.31 & 0.31 \\
        \rowcolor{green!20} gpt4o & BigCodeBench (5-fold) & Reflexion & 5 most effective & 0.30 & 0.35 & 0.35 & 0.36 & 0.36 & 0.36 \\
        gpt4o & BigCodeBench (5-fold) & Reflexion & - & 0.30 & 0.31 & 0.33 & 0.33 & 0.33 & 0.33 \\
        \hline
        \rowcolor{green!20} gpt-o3-mini & BigCodeBench (5-fold) & - & 5 most effective & 0.38 & 0.41 & 0.43 & 0.43 & 0.44 & 0.44 \\
        gpt-o3-mini & BigCodeBench (5-fold) & - & - & 0.38 & 0.42 & 0.42 & 0.42 & 0.42 & 0.42 \\
        \rowcolor{green!20} gpt-o3-mini & BigCodeBench (5-fold) & Reflexion & 5 most effective & 0.39 & 0.41 & 0.44 & 0.46 & 0.48 & 0.48 \\
        gpt-o3-mini & BigCodeBench (5-fold) & Reflexion & - & 0.39 & 0.41 & 0.43 & 0.44 & 0.45 & 0.45 \\
        \hline
    \end{tabular}
    \end{adjustbox}
\end{table*}

In this RQ, we investigate whether our complexity-based feedback can further enhance the performance of feedback-driven code generation agents, specifically Reflexion. To evaluate this, we conducted experiments on BigCodeBench, a more complex dataset compared to HumanEval, MBPP, and LeetCode. We used GPT-4o and GPT-o3 mini as our test models and analyzed their performance with and without Reflexion. The results are presented in Table \ref{tab:tab3}, where Iteration 0 serves as the first baseline (zero-shot generation), while the white rows represent the second baseline (iterative code generation without complexity-based feedback).

Our findings indicate that our complexity-aware feedback method improves code generation performance across iterations, both with and without Reflexion. This is particularly evident in GPT-o3 mini, where iterative complexity-based refinements resulted in noticeable improvements. These results suggest that our approach can be effectively applied on top of agent-based methods, further refining generated code.

However, when comparing the improvements against the second baseline, the differences are relatively minor. This aligns with our findings in RQ1, which indicated that complexity metrics are less predictive of Pass@1 in the BigCodeBench dataset. One possible reason is that BigCodeBench contains more diverse and intricate coding tasks, making it less sensitive to complexity-based refinements.

Nonetheless, the improvements remain slightly higher than the second baseline. This suggests that while complexity-aware feedback may have a limited impact on datasets with weaker complexity-pass@1 correlations, it can still contribute meaningfully when integrated into agent-based code generation workflows.

\mybox{\textbf{Answer to RQ4:} Complexity-based feedback can enhance agent-based code generation. However, improvements were only slightly higher than the second baseline, aligning with our finding that complexity metrics are less predictive of Pass@1 in this dataset. While its impact is limited in lower-accuracy datasets, it can still provide marginal gains in agent-assisted generation.}

%% file: sec/7_Threats.tex
\section{Threats to Validity}



\textbf{Internal Validity:} Fixed parameters (like temperature and max tokens) were set to balance consistency and creativity in code generation. However, broader experimentation with these settings could optimize results for different tasks.

\textbf{External Validity:} While the study uses well-established datasets, these may not fully represent real-world coding challenges. Future work with diverse datasets and newer LLMs could improve generalizability.

\textbf{Construct Validity:} Using SHAP to analyze only the top five metrics might miss other important relationships. Also, relying on LLM-generated test cases could introduce misalignment if they fail to fully capture task requirements, highlighting a need to reduce hallucinations.

\textbf{Conclusion Validity:} The results are robust for the chosen datasets, but expanding the dataset size would improve reliability and generalizability for broader programming scenarios.

%% file: sec/8_Conclusion.tex
\section{Conclusion and Future Work}


This study explored the relationship between code complexity metrics and Pass@1 in LLM-generated code, analyzing their impact on correctness and performance across HumanEval, LeetCode, MBPP, and BigCodeBench. 
Using Logistic Regression models and Shapley values, we identified the most predictive complexity metrics. Prompting LLMs to regenerate code based on these metrics showed that iterative feedback improves Pass@1, refining model outputs over time. Integrating this approach into Reflexion, a feedback-driven code agent, further enhanced code correctness, demonstrating that complexity-aware feedback can complement existing optimization techniques.

Future research can build on these findings by refining LLM-based code generation strategies and exploring broader applications. While this study introduces baseline complexity metrics, developing more sophisticated metrics tailored to machine learning code characteristics could provide more precise feedback and further enhance model accuracy in specialized coding tasks.